# Calculation of the shift exponent for the two layer three state Potts model using the transfer matrix method

Tahmasb Mardani,[1,2] Behrouz Mirza,[2,*] and Mehrdad Ghaemi[3]

[1] *Atomic Energy Organization of Iran, Deputy in Nuclear Fuel Production, Tehran, Iran*
[2] *Department of Physics, Isfahan University of Technology, Isfahan, 84154, Iran*
[3] *Chemistry Department, Teacher Training University, Tehran, Iran*

[*] *E-mail: b.mirza@cc.iut.ac.ir*

## ABSTRACT

A finite-size scaling approach based on the transfer matrix method is developed to calculate the critical temperature and critical exponent of the symmetric and the asymmetric two-layer three-state Potts Models. For similar intralayer interactions our calculation of the shift exponent $\varphi$ confirm some scaling arguments which predict $\varphi = \gamma$, where $\gamma$ is the susceptibility exponent. For unequal intralayer interactions we have obtained $\varphi = 0.5$ which differs from the prediction $\varphi = \gamma/2$ of a generalized mean-field theory.

PACS number: 05.50.+q, 02.70.-c

## I. Introduction

The physical properties of various layered structures and superlattices have been intensely studied both experimentally and theoretically for reasons ranging from fundamental investigations of phase transitions to technical problems encountered in thin-film magnets [1]. Experimentally, submonolayer and monolayer films of ferromagnetic materials offer challenging opportunities to fabricate materials with various magnetic properties, such as giant magnetoresistance, surface magnetic anisotropy, enhanced surface magnetic moment, and surface magnetoelastic coupling. On theoretical grounds, surface magnetism has been treated within several different frameworks: mean-field approximations [2], effective-field theories [3], spin-fluctuation theory [4], renormalization-group methods [5], two-site cluster approximations [6], and Mont Carlo techniques [7]. Though each method has it own advantages, they all have limitations in treating film systems. Numerical techniques such as the Monte Carlo method can provide very accurate results for properties of finite systems; however, they are computation intensive and can be carried out only for relatively small system sizes.

The theoretical works on thin-layer systems are less general. The system of coupled two-dimensional Ising planes on regular lattices, e.g., the square lattice, is not exactly soluble; however, it has been investigated by a variety of approximate methods. Ballentine [8] used high-temperature series expansions to study the model with, $J_1 = J_2 = J_3$, where $J_1$ and $J_2$ are intralayer coupling constants for the first and the second layer, respectively, and $J_3$ is the interlayer coupling constant. This work was later extended by Allan [9] to films up to five layers and by Capehart and Fisher [10] to films up to ten layers. The two-layer system where the interlayer coupling constant differs from the intralayer coupling constant was studied by Abe [11] in the context of a scaling theory valid in the limit of a weak interlayer coupling. The more general case in which $J_1 \neq J_2$ has also received some attention. The most complete treatment was that of Oitmaa and Enting [13], who combined mean-field theory, scaling theory, and high-temperature expansions in a study of the two-layer model, and calculated the variation of the critical temperature, the layer magnetizations, and the interlayer correlation function with $J_3$. They predict the values of the shift exponent $\varphi$, which describes the deviation of the critical temperature $T_C(J_3)$ from the critical temperature in the decoupled limit ($J_3 = 0$),

$$T_C(J_3) - T_C(0) \sim J_3^{1/\varphi} \tag{1}$$

These theories predict that when the coupling $J$ is the same in each layer, then $\varphi = \gamma$, where $\gamma$ is the critical exponent describing the divergence of the susceptibility upon approaching the critical point. Extending these scaling arguments, it is suggested [13] that when the coupling $J$ changes in each sublattice, then $\varphi = \gamma/2$. However, calculations by other methods indicated that the shift exponent for the two-layer Ising model is equal to $1/2$ for different values of the interlayer couplings [14,15,16]. This paper is organized as follows. In Sec. II the transfer matrix method is briefly explained and the shift exponent for the asymmetric two layer Ising model is calculated. In Sec. III, the critical temperatures and the shift exponents for the symmetric and asymmetric two layer three-state Potts models are calculated. A review for the $q$-state Potts models can be found in [20].

## II. Two-Layer Ising Model

Consider a two-layer square lattice with periodic boundary conditions composed of slices, each with two layers with $p$ rows, where each row has $r$ sites. Each slice has then $2 \times p \times r = N$ sites and the coordination number of all sites is the same (namely, 5). In the two-layer Ising model, for any site we define a spin variable $\sigma^{1(2)}(i, j) = \pm 1$, in such a way that $i = 1,...,r$ and $j = 1,...,p$, where the superscript 1(2) denotes the layer number. We include the periodic boundary conditions as

$$\sigma^{1(2)}(i+r, j) = \sigma^{1(2)}(i, j),$$
$$\sigma^{1(2)}(i, j+p) = \sigma^{1(2)}(i, j). \tag{2}$$

In this paper, we discuss the anisotropic ferromagnetic case with nearest neighbor coupling $(J = J_1/kT, J_1 = 2J_2, J_1 = 1)$, where $J_1$ and $J_2$ are the nearest neighbor interactions in the first and second layers, respectively, and with interlayer coupling $J_3/kT$. We take only the interactions among the nearest neighbors into account. The configurational energy for the model may be defined as

$$\frac{E(\sigma)}{kT} = -\frac{1}{kT}\sum_{i=1}^{r^*}\sum_{j=1}^{p^*}\sum_{n=1}^{2}\{J_n\sigma^n(i,j)\sigma^n(i+1,j)+J_n\sigma^n(i,j)\sigma^n(i,j+1)\}$$

$$-\frac{J_3}{kT}\sum_{i=1}^{r}\sum_{j=1}^{p}\sigma^1(i,j)\sigma^2(i,j) \quad (3)$$

where the asterisk indicates the periodic boundary conditions (Eqs.2). The canonical partition function $Z(J)$ is

$$Z(J) = \sum_{\{\sigma\}} e^{\frac{-E(\sigma)}{kT}} \quad (4)$$

Substitution of Eq. (3) into Eq. (4) gives,

$$Z(J) = \sum_{\sigma(\{i\},1)} \cdots \sum_{\sigma(\{i\},p)} \langle 1|T|2\rangle\langle 2|T|3\rangle \ldots \langle p|T|1\rangle \quad (5)$$

where

$$|j\rangle = |\sigma^1(1,j)\rangle \otimes |\sigma^2(1,j)\rangle \otimes |\sigma^1(2,j)\rangle \otimes |\sigma^2(2,j)\rangle \ldots \otimes |\sigma^2(r,j)\rangle \quad (6)$$

$$\sum_{\sigma(\{i\},j)} = \sum_{\sigma^1(1,j)}\sum_{\sigma^1(2,j)}\cdots\sum_{\sigma^1(r,j)}\sum_{\sigma^2(1,j)}\sum_{\sigma^2(2,j)}\cdots\sum_{\sigma^2(r,j)} \quad (7)$$

By orthogonal transformation, the **T** matrix can be diagonalized where Eq. (4) for the large values of $p$ can be written as

$$Z(J) = tr\, T^p \sim (\lambda_{max})^p, \quad (8)$$

where $\lambda_{max}$ is the largest eigenvalue of **T**. From the well known thermodynamic relation for the Helmholtz free energy, $A = -kT \ln Z$, along with Eq. (8) the following results are obtained:

$$a(J) = \frac{-A}{NkT} = \frac{\ln \lambda_{max}}{r}, \quad (9)$$

$$u(J) = \frac{\partial a(J)}{\partial J} \quad (10)$$

where $u(J)$ and $a(J)$ are the reduced internal energy and Helmholtz free energy per site, respectively.

For the two-layer square lattice with size $r$, using Eq. (8), the elements of the ***T*** matrix have been calculated numerically. We have employed a method [17,18,19] for reducing the size of the transfer matrix and the $\lambda_{\max}$ was calculated with a high precision for different $J_3$ values. The reduced internal energy has been calculated for a two-layer square lattice with $r = 3,4,5,6,7,8$ for different values of $J_3$. In order to obtain the value of the critical temperature $J_C$, the intersection point for two unlimited lattices with different sizes should be found [19]. However, such a point may be predicted if we have an expression for the intersection point in terms of $1/n$, where $n = 4rr'$. As shown in an earlier work, for the two dimensional and the two-layer Ising model the critical temperature may be approximated as a general polynomial of a certain degree,

$$J_n = \sum_j a_j \left(\frac{1}{n}\right)^j \qquad (11)$$

where the values of the $a_j$'s have been calculated using the least square method [19]. If we assume that Eq. (11) is applicable for large lattice sizes then for the limit of $n \to \infty$, $J_C$ is equal to $a_0$. We have done similar calculations for the two dimensional three-state Potts model [21]. Our results were in complete agreement with earlier results already obtained by other methods. So one may decide to generalize the method of calculation to the two-layer Potts model. Such calculations have been done for different values of $J_3$. The results are given in Table 1. In Fig. 1 we illustrated the intersection and its extrapolated critical points. It is clear that when the size of the lattice becomes larger the intersection points approach each other and so the critical point can be investigated.

In this model an interesting situation appears when the interlayer coupling $J_3$ becomes infinitesimally small compared to the interlayer coupling $J_1$ and $J_2$. For this case we have used Eq. 1 and data in Table 1 to obtain the shift exponent $\varphi$. The calculated values for this parameter are given in Table 2. Our calculation for the shift exponent of the asymmetric two-layer Ising model gives $\varphi \sim 0.55$ and is in agreement with the result which has already been obtained in [15].

### III. Two-Layer three-state Potts Model

Although we do not know the exact solution of the Potts model for two dimensional lattices at present time, a large amount of numerical information has been accumulated for the critical properties of the various Potts models. The reason for the extension of our approach to the two-layer three-state Potts model is the fact that such a model is an important testing ground for different methods and approaches in the study of critical phenomena.

For the two-layer square lattice Potts model with size *r*, we can use transfer matrices as in the two-layer Ising model and calculate the largest eigenvalues. The size of the original matrix for the two-layer three-state Potts model is much larger than the size of the matrix of the two-layer Ising model and so numerical calculations are more difficult. For example, on strips with width $r = 3, 4$ we have matrices of order 64, 254 for the two-layer Ising model, and matrices of order 729, 6561 for the two-layer three-state Potts model, respectively. The order of transfer matrices for strips with $r, r' = 2, 3, 4,$ and 5 is equal to 81, 729, 6561, and 59049 which can be reduced to 21, 92, 498, and 3210, respectively. The size of the reduced transfer matrix is much smaller than that of the original transfer matrix $T$ and so $\lambda_{\max}$ can be easily calculated from the reduced matrix.

In this section we first investigate the three-state Potts model on a symmetric two layer lattice, where the intralayer couplings are the same $(J_1 = J_2)$. By increasing the interlayer couplings $J_3$ from 0 to 1.2 we have calculated the reduced free energy per site for four different strips with $r, r' = 2, 3, 4, 5$. We have calculated the critical points by a similar method which is explained in Sec. II. The critical points and shift exponent are given in Tables 3 and 4. The value of the shift exponent is equal to $1.44 \pm 0.05$.

For a symmetric two-layer lattice the scaling theories [11,12] predict that when the coupling $J$ is the same in each layer, then $\varphi = \gamma$, where $\gamma$ is the critical exponent describing the divergence of susceptibility upon approaching the critical point. Our result is in agreement with the theoretical value, i.e., $\varphi = 1.44$. However, for asymmetric two-layer models this theoretical prediction is not correct. For an asymmetric two-layer Ising model the value of the shift exponent is equal to $1/2$ [15].

Our aim in this work is to calculate the shift exponent for the three-state Potts model on an asymmetric two-layer lattice. Our numerical calculations for the critical points and the shift exponent are given in Tables 5 and 6. It is interesting that the shift exponent is equal to $1/2$ and its value is equal to the Ising model case. In this part we argue that one may expect the value $1/2$ for the shift exponent.

Let us consider four spins in the corners of an elementary square with interactions like those shown in Fig. 2. Decimating over $s_3$ and $s_4$ (spins of the second layer), we obtain that the coupling between $s_1$ and $s_2$ changes by $J_\delta$. This additional interaction has to satisfy the condition

$$Ae^{\beta J_\delta [\delta(s_1, s_2)]} = \sum_{s_3, s_4} e^{\beta \{J_3 [\delta(s_1, s_3) + \delta(s_2, s_4)] + J_2 \delta(s_3, s_4)\}} \tag{12}$$

where *A* is an unimportant factor. This condition is equivalent to the following set of equations:

$$\begin{aligned} Ae^{\beta J_\delta} &= e^{\beta(2J_3 + J_2)} + 4e^{\beta J_3} + 2e^{\beta J_2} + 2, \\ A &= 2e^{\beta(J_3 + J_2)} + e^{2\beta J_3} + 2e^{\beta J_3} + 2e^{\beta J_2} + 3, \end{aligned} \tag{13}$$

with the solution

$$J_\delta = \frac{1}{\beta} \ln \left[ \frac{e^{\beta(2J_3 + J_2)} + 4e^{\beta J_3} + 2e^{\beta J_2} + 2}{2e^{\beta(J_3 + J_2)} + e^{2\beta J_3} + 2e^{\beta J_3} + 2e^{\beta J_2} + 3} \right]. \tag{14}$$

Expanding the right-hand side of Eq. (14) one can easily check that for small $J_3$ we have $J_\delta \sim J_3^2$. Thus, spins of the second layer effectively change the coupling in the first layer into $J_1' = 1 + CJ_3^2$, where $C$ is a certain constant. Since the critical temperature of the square lattice Potts model is proportional to the coupling, thus $T_C(J_3) - T_C(0) \sim J_3^2$ and $\phi = 1/2$ easily follows. At this time we do not know a complete proof for our numerical result. The calculated value is equal to $1/2$ which is different from that of the original scaling arguments [11,12]. It seems that for unequal intralayer interactions the shift exponent is equal to $1/2$ for an arbitrary $q$-state Potts model.

## IV. Scaling analysis

One may use a phenomenological renormalization based on the largest eigenvalue $\lambda_{max}$ of the transfer matrix to verify the above arguments [22]. According to finite size scaling theory, the singular part of the free energy per spin of the s spins has the scaling from $a_r(t,J_3) = r^{-d} A(r^y t, r^{\varphi y} J_3)$, where $t \propto T - T_C(J_3 = 0)$, $d = 2$, and $y = 1/\nu$ is the thermal scaling index ($a_r = \ln \lambda_{max}/r$). For the Ising model $y = 1$, and for the three-state Potts model $y = 6/5$. Differentiating $a_r(t,J_3)$ twice with respect to $t$, one finds that the specific heat per spin scales as $c_r(t,J_3) = r^{2y-d} G(r^y t, r^{\varphi y} J_3)$ and so for $J_3 = 0$ we obtained the thermal exponent $y$ from

$$\left(\frac{r'}{r}\right)^{d-2y} = \frac{c_r}{c_{r'}} \qquad (15)$$

Differentiating $a_r(t,J_3)$ twice with respect to $J_3$ we arrive at

$$\left(\frac{r'}{r}\right)^{d-2\varphi y} = \frac{\partial^2 a_r / \partial J_3^2}{\partial^2 a_{r'} / \partial J_3^2} \qquad (16)$$

and so one may obtained the shift exponent by using Eq. (16). We have obtained the thermal exponent $y$ by evaluating the right-hand side of Eq. (15) for *t=0*, i.e., $J_C = \ln(1+\sqrt{3})$ and $J_3 = 0$, where the results are given in Table 7. Then $\varphi y$ is calculated by considering the right-hand side of Eq. (16) for *t=0* i.e., $J_3 = 0$ [small values of $J_3$ have been used for differentiating $a_r(t,J_3)$ twice with respect to $J_3$]. Finally, from the values of *y* and $\varphi y$ one can find the shift exponent which is given in Table 7. For the asymmetric two-layer three-state Potts model with different lattice sizes $r, r' = 2,3,4,5$, the calculated shift exponent by this method is about 0.6. A part of our results for $J_3 = 0.02, 0.03, 0.05, 0.07, 0.09$ are given in Table 7.

## V. Conclusion

In this paper a numerical method has been used to calculate the critical properties for the two-layer spin systems. The critical temperatures and the shift exponent are calculated. For the symmetric two-layer three-state Potts model with similar intralayer interactions our calculations of the shift exponent $\varphi$ confirm some scaling arguments which predict $\varphi = \gamma = 1.44$. For the asymmetric two-layer three-state Potts model with unequal intralayer interactions our calculations of the shift exponent $\varphi$ differs from the prediction $\phi = \gamma/2$ of a generalized mean-field theory. Our result indicates that the shift exponent for the two-state Ising and the three-state Potts models are equal to $1/2$. This result implies that the value of the shift exponent may be independent of the value of $q$ in the $q$-state Potts model.

## Acknowledgements

Our thanks go to the Isfahan University of Technology and the Atomic Energy Organization of Iran for the financial support they made available to us.

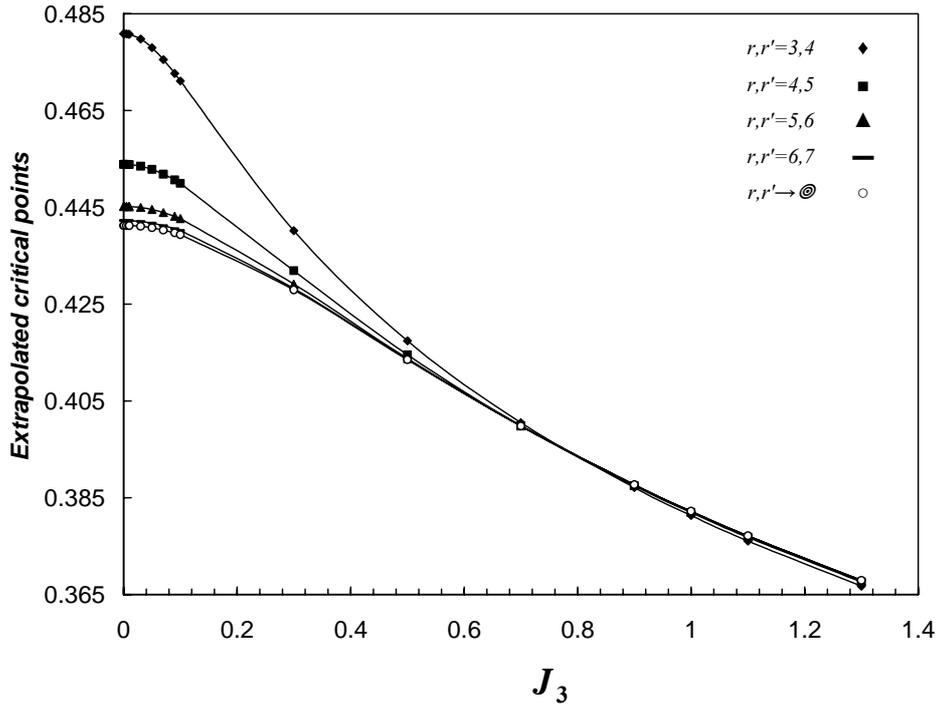

FIG. 1: Extrapolated critical temperature for the asymmetric two-layer Ising model. Curves with black points include intersection points and curve with empty circles includes extrapolated critical temperatures.

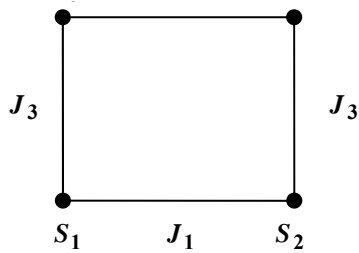

FIG. 2: An elementary interlayer square. We decimate over spins $s_3$ and $s_4$.

TABLE.1: The calculated critical temperature of the asymmetric two-layer Ising model [the exact value for $J_3 = 0$ is $J_c = (1/2)\ln(1+\sqrt{2}) = 0.440687$].

| $J_3$ | Calculated critical temperature $(J_C)$ |
|---|---|
| 0 | 0.439846 |
| 0.001 | 0.439845 |
| 0.005 | 0.439837 |
| 0.009 | 0.439822 |
| 0.01 | 0.439815 |
| 0.03 | 0.439595 |
| 0.05 | 0.439182 |
| 0.07 | 0.438625 |
| 0.09 | 0.437968 |
| 0.1 | 0.437607 |
| 0.3 | 0.427430 |
| 0.5 | 0.413695 |
| 0.7 | 0.400044 |
| 0.9 | 0.387811 |
| 1 | 0.382240 |
| 1.1 | 0.377145 |
| 1.3 | 0.367875 |

TABLE. 2: Shift exponent for the asymmetric two-layer Ising model [the exact value for $J_3 = 0$ is expected to be $\varphi = 1/2$].

| $J_3$ | $\varphi$ |
|---|---|
| 0.07,0.09,0.1,0.3 | 0.559653 |
| 0.05,0.07,0.09,0.1,0.3 | 0.576416 |
| 0.05,0.07,0.09,0.1 | 0.584866 |

TABLE. 3: The calculated critical temperature of the symmetric two-layer three-state Potts model [the exact value for $J_3 = 0$ is $J_c = \ln(1+\sqrt{3}) = 1.005052$].

| $J_3$ | Calculated critical temperature $(J_C)$ |
|---|---|
| 0.000 | 1.052495 |
| 0.003 | 0.982836 |
| 0.005 | 0.980552 |
| 0.007 | 0.978269 |
| 0.009 | 0.975994 |
| 0.010 | 0.974860 |
| 0.050 | 0.935379 |
| 0.090 | 0.907997 |
| 0.100 | 0.902499 |
| 0.500 | 0.795385 |
| 0.600 | 0.778996 |
| 0.700 | 0.764166 |
| 0.800 | 0.750573 |
| 0.900 | 0.737999 |
| 1.000 | 0.726306 |
| 1.100 | 0.715379 |
| 1.200 | 0.705136 |

TABLE. 4: Shift exponent for the symmetric two-layer three-state Potts model [the exact value for $J_3 = 0$ is $\varphi = \gamma = 13/9 = 1.444444$].

| $J_3$ | $\varphi$ |
|---|---|
| 0.003,0.005,0.007 | 1.45285 |
| 0.003,0.005,0.007,0.009 | 1.42794 |
| 0.003,0.005,0.007,0.009,0.01,0.05 | 1.38958 |
| 0.003,0.005,0.007,0.009,0.01,0.05,0.09, 0.1,0.5 | 1.35199 |

TABLE. 5: The calculated critical temperature of the asymmetric two-layer three-state Potts model [the exact value for $J_3 = 0$ is $J_c = \ln(1+\sqrt{3}) = 1.005052$].

| $J_3$ | Calculated critical temperature $(J_C)$ |
|---|---|
| 0.000 | 1.052495 |
| 0.001 | 1.050030 |
| 0.003 | 1.045497 |
| 0.005 | 1.041418 |
| 0.007 | 1.037615 |
| 0.009 | 1.034370 |
| 0.010 | 1.032804 |
| 0.030 | 1.020268 |
| 0.050 | 1.011642 |
| 0.070 | 1.000748 |
| 0.090 | 0.994273 |
| 0.100 | 0.989953 |
| 0.500 | 0.943733 |
| 0.600 | 0.931359 |
| 0.700 | 0.918790 |
| 0.800 | 0.906272 |
| 0.900 | 0.893978 |
| 1.000 | 0.882024 |
| 1.100 | 0.870482 |
| 1.200 | 0.859397 |

TABLE. 6: Shift exponent for the asymmetric two-layer three-state Potts model [the exact value for $J_3 = 0$ is expected to be $\varphi = 1/2$].

| $J_3$ | $\varphi$ |
|---|---|
| 0.07,0.09,0.1 | 0.518406 |
| 0.07,0.09,0.1,0.5,0.6,0.7 | 0.487023 |
| 0.07,0.09,0.1,0.5 | 0.504776 |
| 0.07,0.09,0.1,0.5,0.6 | 0.494463 |

TABLE. 7: Finite-size scaling results for thermal and shift exponents of the asymmetric two-layer three-state Potts model.

| $r/r'$ | $J_3$ | $y$ | $\varphi$ |
|---|---|---|---|
| 4/6 | 0.02,0.03,0.05 | 1.292775 | 0.749020 |
| 6/8 | | 1.28093 | 0.707055 |
| 8/10 | | 1.25435 | 0.615987 |
| 4/6 | 0.07,0.09,0.1 | 1.29278 | 0.725818 |
| 6/8 | | 1.31703 | 0.681638 |
| 8/10 | | 1.31783 | 0.650203 |
| Exact value | | 6/5 | 1/2 |